\documentclass[twocolumn]{aastex61}

\pdfoutput=1

\received{July 18, 2017}
\revised{October 24, 2017}
\accepted{November 14, 2017}
\submitjournal{ApJ}

\shorttitle{LMXBs in the outer halo of NGC 4472}
\shortauthors{van~Haaften et al.}

\begin{document}

\title{An excess of low-mass X-ray binaries in the outer halo of NGC 4472}

\correspondingauthor{Lennart~M.~van~Haaften}
\email{L.vanHaaften@ttu.edu}

\author[0000-0001-5422-6775]{Lennart~M.~van~Haaften}
\affiliation{Department of Physics and Astronomy, Texas Tech University, Box 41051, Lubbock, TX 79409-1051, USA}

\author{Thomas~J.~Maccarone}
\affiliation{Department of Physics and Astronomy, Texas Tech University, Box 41051, Lubbock, TX 79409-1051, USA}

\author{Paul~H.~Sell}
\affiliation{Physics Department, University of Crete, 71003 Heraklion, Crete, Greece}
\affiliation{Department of Physics and Astronomy, Texas Tech University, Box 41051, Lubbock, TX 79409-1051, USA}

\author{J.~Christopher~Mihos}
\affiliation{Department of Astronomy, Case Western Reserve University, 10900 Euclid Ave, Cleveland, OH 44106, USA}

\author{David~J.~Sand}
\affiliation{Department of Physics and Astronomy, Texas Tech University, Box 41051, Lubbock, TX 79409-1051, USA}
\affiliation{Steward Observatory, The University of Arizona, 933 N. Cherry Avenue, Tucson, AZ 85721, USA}

\author{Arunav~Kundu}
\affiliation{Department of Physics and Astronomy, Texas Tech University, Box 41051, Lubbock, TX 79409-1051, USA}
\affiliation{Eureka Scientific, Inc., 2452 Delmer Street, Suite 100, Oakland, CA 94602, USA}
\affiliation{Computational Physics, Inc., 8001 Braddock Road, Suite 210, Springfield, VA 22151, USA}

\author{Stephen~E.~Zepf}
\affiliation{Department of Physics and Astronomy, Michigan State University, East Lansing, MI 48824, USA}

\begin{abstract}

We present new \textit{Chandra} observations of the outer halo of the giant elliptical galaxy NGC 4472 (M49) in the Virgo Cluster. The data extend to $130$ kpc ($28\arcmin$), and have a combined exposure time of $150$ ks.
After identifying optical counterparts using the Next Generation Virgo Cluster Survey to remove background active galactic nuclei and globular cluster (GC) sources, and correcting for completeness, we find that the number of field low-mass X-ray binaries (LMXBs) per unit stellar \textit{V}-band light increases significantly with galactocentric radius.
Because the flux limit of the complete sample corresponds to the Eddington limit for neutron stars in NGC 4472, many of the ${\sim}90$ field LMXBs in this sample could host black holes.
The excess of field LMXBs at large galactocentric radii may be partially caused by natal kicks on black holes and neutron stars in binary systems in the inner part of the galaxy.
Furthermore, since the metallicity in the halo of NGC 4472 strongly decreases towards larger galactocentric radii, the number of field LMXBs per unit stellar mass is anti-correlated with metallicity, opposite to what is observed in GCs. Another way to explain the spatial distribution of field LMXBs is therefore a reversed metallicity effect, although we have not identified a mechanism to explain this in terms of stellar and binary evolution.

\end{abstract}


\section{Introduction}
\label{sect:halo:intro}

X-ray observations of old stellar populations such as those of elliptical galaxies, spiral galaxy bulges, and globular clusters (GCs) reveal numerous, bright X-ray point sources. These are almost exclusively low-mass X-ray binaries (LMXBs) \citep{trinchieri1985}, stellar systems in which a neutron star (NS) or black hole (BH) accretes matter from a low-mass (mass \mbox{$\lesssim 1.5\ M_{\sun}$}) companion star overfilling its Roche lobe. Their population sizes and properties are important for our understanding of stellar and binary evolution \citep[e.g.,][]{tauris2006}.

The formation rate of LMXBs depends on many factors, the two most important being the total stellar mass and the number density of stars, both of which are clearly demonstrated by observations. The number of LMXBs is positively correlated with the total stellar mass, since low-mass stars contain the majority of the stellar mass in a galaxy \citep[e.g.,][]{gilfanov2004}. The number of LMXBs is also positively correlated with the stellar density, even per unit stellar mass. Many LMXBs are found in GCs, which contain about two orders of magnitude more LMXBs per unit mass than in the field \citep{katz1975}. In early-type galaxies, roughly half of LMXBs reside in GCs \citep[e.g.,][]{sarazin2000,angelini2001,kundu2002}. This can be explained by the high frequency of dynamical interactions in these stellar systems \citep[e.g.,][]{clark1975}.

These and other factors affect the spatial distribution of LMXBs in a galaxy. While the distribution of LMXBs is strongly affected by the distributions of the stellar mass and GCs, varying metallicity and migration caused by supernova kicks and could alter it as well.

\subsection{Kicks received by black holes and neutron stars}

Natal kicks of NSs generally have velocities of several hundreds of km s$^{-1}$ \citep{lyne1994}.
There is also evidence that BHs receive high natal kick velocities, close to those of NSs \citep[e.g.,][]{brandt1995,fragos2009,repetto2015}, as well as evidence for misaligned spins caused by asymmetric kicks \citep[e.g.,][]{hjellming1995,ingram2009}.

However, some NSs \citep{pfahl2002,podsiadlowski2004,vandenheuvel2007} and BHs \citep[e.g.,][]{millerjones2009,reid2014} are known to receive weak kicks.
Overall, the BH kick velocity distribution is not as well constrained as that of NSs and current evidence points to a fairly wide range of BH kick velocities.

\subsection{The metallicity effect for LMXBs}

The \textit{metallicity effect} is the observation that metal-rich GCs have a higher chance to host an LMXB than metal-poor clusters. High metallicity is indicated by late collective spectral type and low galactocentric distance.

\citet{kundu2002} established the metallicity effect using \textit{Chandra X-ray Observatory} data of NGC 4472, finding that metal-rich (red) GCs are about $2.7$ times more likely to host an LMXB than metal-poor (blue) clusters, and also showed clearly that the effect was not just due to a possible dependence of color on galactocentric radius or cluster luminosity \citep[i.e., stellar population size, see also][]{sarazin2003}.

\subsubsection{Proposed explanations for the metallicity effect}

If metal-rich environments would produce increased numbers of massive stars, or stars with large radii, more X-ray binaries would be expected to form, possibly explaining the metallicity effect.

\citet{grindlay1987,grindlay1993} argued that a flatter \textit{initial} mass function implies a higher number of NSs, and hence more LMXBs, possibly (partially) explaining the metallicity effect.

\citet{bellazzini1995} argued for a dependence on metallicity of stellar radii as well as stellar masses, which would increase tidal capture rates and fill Roche lobes at larger orbital separations.
\citet{ivanova2012} argued that in metal-rich GCs, red giants are more massive and more numerous than in metal-poor ones. This leads to a higher collision and binary exchange frequencies.

Other explanations involve the timescale of binary evolution.
\citet{iben1997} and \citet{maccarone2004} pointed out that metal-poor donors have less line cooling and therefore stronger irradiation-induced stellar winds, and that as a result these binaries lose more angular momentum, evolve faster, and live shorter.

Another resolution was proposed by \citet{ivanova2006}, who found that metal-rich donor stars in the mass range $0.85 - 1.25\ M_{\sun}$ have an outer convection zone, and as a result can experience magnetic braking, which is an important source of angular momentum loss that drives binary evolution. Stars with masses in the upper part of this range may no longer exist in old populations, but their descendants can still be X-ray binaries with lower donor masses.

\subsection{Aim of this work}

Some of the models for the metallicity effect described in the previous section, such as increased stellar winds and outer convection zones, have the same effect regardless of stellar interaction rates, and therefore predict the same metallicity effect in GCs and the field.
The remaining mechanisms work in dynamical environments for additional reasons than in low-density environments, and predict a stronger metallicity effect in GCs compared to the field. More NSs, larger stellar radii, and higher stellar masses result in more captures in GCs, while they also result in, respectively, more primordial LMXBs, more easily filled Roche lobes, and stronger magnetic braking in environments of any stellar density.

In order to disentangle the cause of the metallicity effect, one can test whether the metallicity effect is the same or different in an environment where dynamical interactions between stars and binaries do not contribute to the formation of LMXBs, i.e., the field. If the metallicity effect is the same in the field, then increased stellar winds and outer convection zones may be the best explanation of the metallicity effect. If it is weaker in the field, then the other explanations are more likely. If the effect is stronger in the field, further explanations are needed, such as natal kicks, the effect of which we also explore.

Here we present such a study of the metallicity effect in the field using new \textit{Chandra} data of the outer halo of the giant elliptical galaxy \object{NGC 4472} (M49), a member of the \object{Virgo Cluster}. The NGC 4472 halo is massive and extended, and \citet{mihos2013} determined that the halo has steep color and metallicity gradients between $30 - 100$ kpc, making it an excellent environment to test the metallicity effect. Our \textit{Chandra} observations cover a larger angular radius and resolution than those obtained so far for any galaxy. We also used archival \textit{Chandra} data of the inner region of NGC 4472 to complement the new data. We determined the number of field LMXBs per unit stellar light for a range of galactocentric radii, and hence a range of inferred metallicities.

\section{Observations and data analysis}
\label{sect:halo:obs}

We adopted a distance of $16.3$ Mpc to NGC 4472 \citep{tonry2001}, which is used to convert flux rates to luminosities.
The center of the galaxy is located at R.A. 12$^{\mathrm{h}}$ 29$^{\mathrm{m}}$ 46\fs80, Dec +08\degr 00\arcmin 01\farcs48 (Chandra X-ray Center Data Archive).

\subsection{\textit{Chandra} X-ray observations}

\begin{table*}
\caption{List of \textit{Chandra} observations of NGC 4472.}
\label{tab:obs}
\begin{tabular}{ccccccc}
\hline
Observation ID & Instrument & Exposure time (ks) & Data mode & RA & Dec & Start date and time \\
\hline
321 & ACIS-S3 & 39.59 & Very faint & 12 29 46.90 & +08 00 13.00 & 2000 Jun 12 01:47:47 \\
15756 & ACIS-I & 32.07 & Faint & 12 29 04.40 & +07 49 24.10 & 2014 Apr 16 09:50:28 \\
15757 & ACIS-I & 29.68 & Faint & 12 28 45.00 & +08 03 48.00 & 2014 Apr 18 00:01:11 \\
15758 & ACIS-I & 29.67 & Faint & 12 29 43.20 & +08 15 12.30 & 2014 Apr 20 13:52:10 \\
15759 & ACIS-I & 29.68 & Faint & 12 30 43.80 & +08 05 21.60 & 2014 Apr 25 18:45:47 \\
15760 & ACIS-I & 29.38 & Faint & 12 30 14.20 & +07 50 00.20 & 2014 Apr 26 03:26:34 \\
\hline
\end{tabular}
\end{table*}

\begin{figure}
\plotone{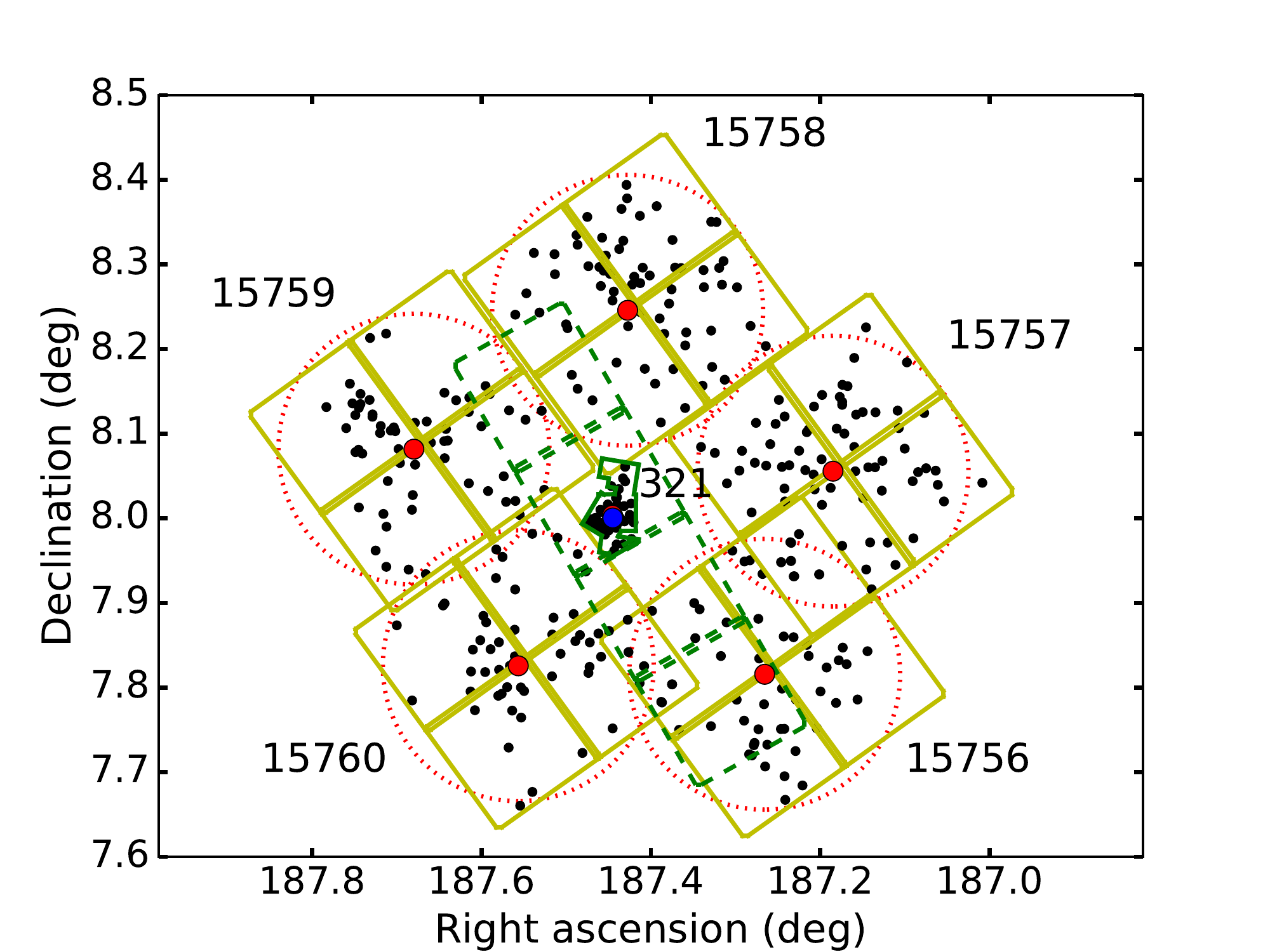} % Chandra_data/NGC4472/halo
\caption{\textit{Chandra} field of view of NGC 4472 used in this study. Yellow fields are the five new ACIS-I observations consisting of four CCDs each, with aimpoints shown as red circles. The archival ACIS-S field of the inner region is shown as dashed green, with the \textit{HST} fields within the ACIS-S3 field in solid green. \textit{Chandra} Observation IDs are shown in or next to the corresponding fields. The blue circle marks the center of the galaxy. The red dotted circles around each ACIS-I aimpoint depict the $9\farcm6$ radial limits used to obtain a flux-limited complete sample (Sect.~\ref{sect:halo:completeness}). Black dots indicate all X-ray sources identified by \textsc{wavdetect}, including field and GC LMXBs, and background AGN.}
\label{fig:halo:fov}
\end{figure}

We obtained five new Advanced CCD Imaging Spectrometer I (ACIS-I) observations covering a large fraction of the halo of NGC 4472, with small overlaps between the fields (Table~\ref{tab:obs} and Fig.~\ref{fig:halo:fov}). The galactocentric radii of these fields range from $3\farcm2 - 28\farcm4$ ($15 - 135$ kpc, or $1 - 9$ effective radii, using $r_\mathrm{e} = 3\farcm24 \pm 0\farcm28$ \citep{kormendy2009}).
We also used an archival ACIS-S observation (ObsID 321) of the inner region, which is not covered by the new observations.

Sources in the \textit{Chandra} observations were detected using the CIAO tool \textsc{wavdetect} \citep{freeman2002}, with the \textsc{sigthresh} parameter set at $10^{-6}$, and the \textsc{scales} values are a geometric progression with a constant factor of $\sqrt{2}$ between $2$ and $16$.
Fluxes and their uncertainties were determined with the CIAO tool \textsc{srcflux} with an absorbed power law model with a photon index of $1.7$ and a Galactic foreground neutral hydrogen column density based on the model by \citet{dickey1990}. The average value in the direction of NGC 4472 is $N_\mathrm{H} \approx 1.6 \times 10^{20}$ cm$^{-2}$, according to the HEASARC W3 $N_\mathrm{H}$ tool based on \citet{dickey1990,kalberla2005}.

We divided the galaxy into $9$ elliptical annuli on the sky with a constant width of $2\farcm63$ along the major axis, centered around the center of NGC 4472. The position angle of the major axis is $-31\degr$ ($31\degr$ west of north), and constant ellipticity $1-b/a = 0.28$, with $b/a$ the ratio between the lengths of the minor and major axes, following the shape of the isophotes used in \citet{mihos2013} based on \citet{kormendy2009,janowiecki2010}. For each X-ray source we determined in which annulus it is located. A~higher number of annuli would make Poisson errors on the number of X-ray sources per annulus larger than is practical.

Because star formation in NGC 4472 is expected to have ended at least several Gyr ago \citep{thomas2005,baes2007}, stars with masses over $1.5\ M_{\sun}$ will no longer exist, so high-mass X-ray binaries are not expected in this field.
That means that X-ray sources brighter than the sensitivity limit are either LMXBs or active galactic nuclei (AGN).
LMXBs in GCs and background AGN are still included in Figs.~\ref{fig:halo:fov}--\ref{fig:halo:options}, but will be statistically removed per annulus using the method described in Sect.~\ref{sect:halo:matching}.

\subsubsection{Completeness limits}
\label{sect:halo:completeness}

\begin{figure}
\plotone{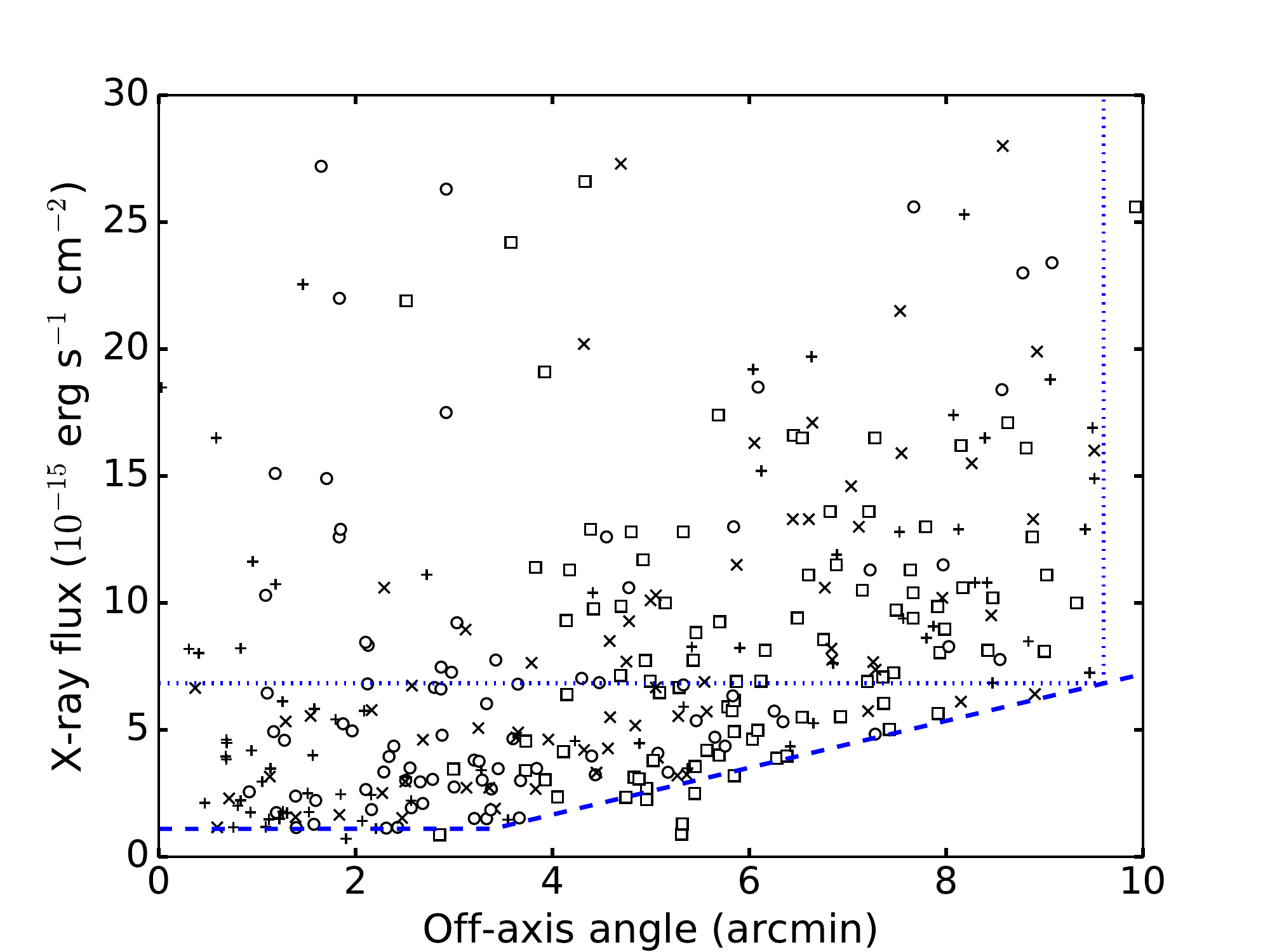} % Chandra_data/NGC4472/halo
\caption{Full sample of $344$ X-ray sources from the six observations listed in Table~\ref{tab:obs}, with flux in the $0.5-7$ keV band plotted against angular distance from the respective aimpoints. Plus symbols indicate sources with galactocentric radii under $10\arcmin$, crosses between $10\arcmin - 14\arcmin$, circles between $14\arcmin - 18\arcmin$, and squares above $18\arcmin$. The dashed blue line, consisting of a horizontal segment at $1.1 \times 10^{-15}$ erg s$^{-1}$ cm$^{-2}$ and a linear increasing segment, approximates the sensitivity limit as a function of off-axis angle. The dotted blue box (the region above $6.8 \times 10^{-15}$ erg s$^{-1}$ cm$^{-2}$ and below $9\farcm6$) contains the complete sample used in this study. $32$ sources lie outside the plotted parameter space.}
\label{fig:halo:oaa}
\end{figure}

Further away from the aimpoint on each chip, the point spread function becomes larger and the sensitivity worse. Figure~\ref{fig:halo:oaa} shows the fluxes of all sources against the off-axis angle from the aimpoint of their respective observations (which is different from galactocentric radius). In order to have a flux-limited complete sample, we discarded all sources fainter than a certain chosen flux limit, and also discarded sources located outside the off-axis angle corresponding to this flux limit (dashed blue line in Fig.~\ref{fig:halo:oaa}).

\begin{figure}
\plotone{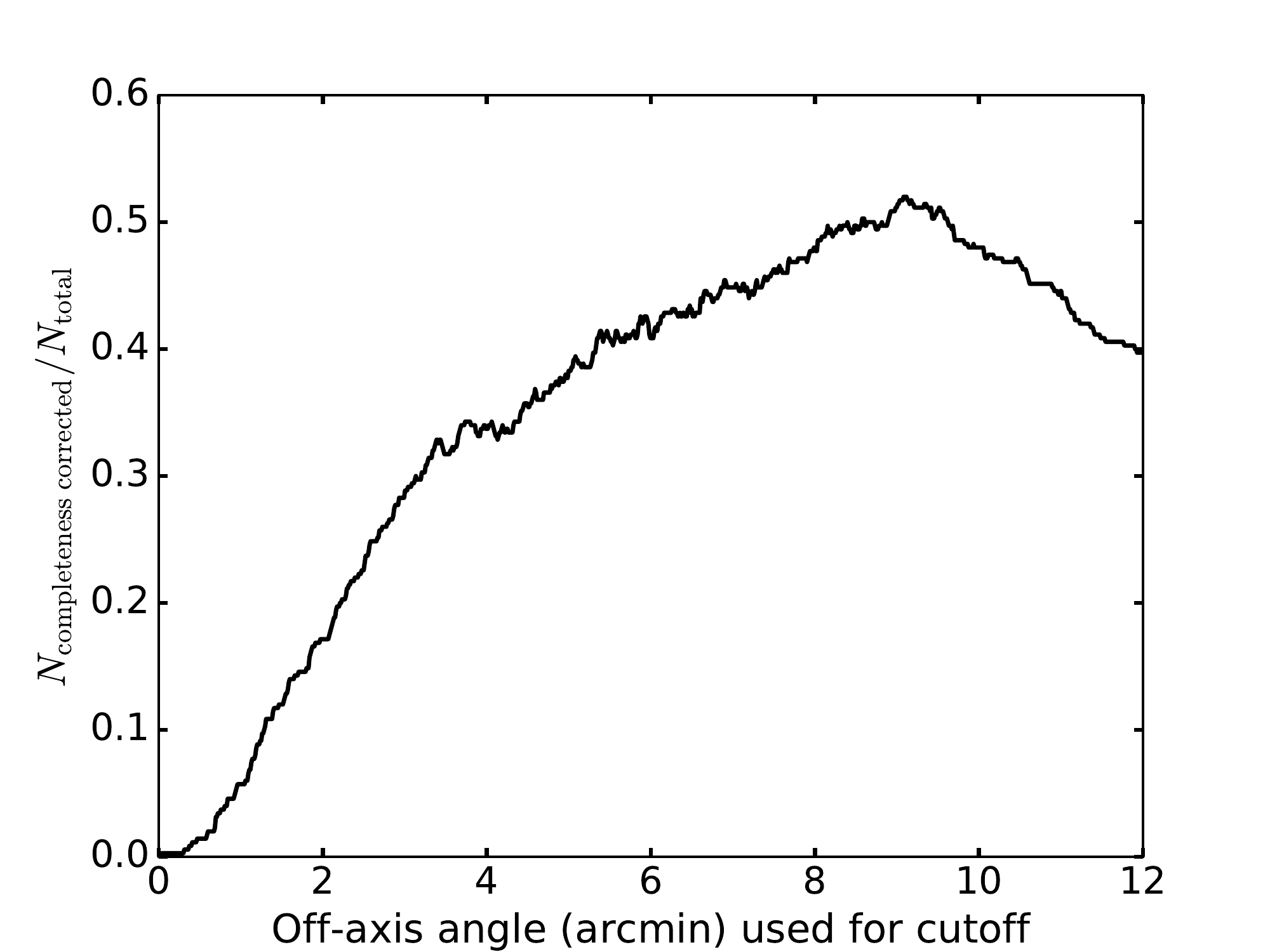} % Chandra_data/NGC4472/halo
\caption{Fraction of $N_\mathrm{total} = 344$ sources that are included in the flux-limited complete sample for a chosen off-axis angle cutoff, using the dashed blue line in Fig.~\ref{fig:halo:oaa}.}
\label{fig:halo:options}
\end{figure}

Choosing a low limiting flux means that more faint sources are included, but also that many sources far away from their respective aimpoints are discarded. A higher flux limit has the opposite effects, and in between is an optimum where over $50\%$ of sources are retained.
The fraction of all $344$ sources identified by \textsc{wavdetect} that are included in a flux-limited complete sample (i.e., that is brighter than the flux limit and located within the corresponding off-axis radius) as a function of chosen off-axis angle cutoff is visualized in Fig.~\ref{fig:halo:options}. In this study, we adopted an angle of $9\farcm6$, at the maximum in the curve, corresponding to a completeness flux limit of $6.8 \times 10^{-15}$ erg s$^{-1}$ cm$^{-2}$ for the $0.5-7$ keV band (${\sim} 16$ net counts for the faintest sources in the new halo observations). This converts to $2 \times 10^{38}$ erg s$^{-1}$ for sources located at the distance of NGC 4472. At the end of Sect.~\ref{sect:halo:results}, we present results for a smaller angle of $5\farcm4$ and a larger angle of $12\arcmin$.

The aimpoints of the five ACIS-I fields are located at $12\farcm4, 14\farcm8, 14\farcm8, 15\farcm4,$ and $15\farcm8$ from the center of the galaxy. The variation in these distances allows for a larger coverage of galactocentric radii by a complete sample for a given off-axis angle cutoff.
Due to the overlap between fields, six sources are detected in two observations. For these sources, we used the average flux and show the source at the smallest off-axis angle in Fig.~\ref{fig:halo:oaa}.

\subsection{Optical data}
\label{sect:opt}

To distinguish field LMXBs from AGN, LMXBs in GCs, and foreground stars, we used deep optical observations from the Next Generation Virgo Cluster Survey point-source catalog \citep[NGVS,][]{ferrarese2012} taken with MegaCam on the Canada--France--Hawaii Telescope. This survey detected $180\,692$ optical sources with a depth of apparent magnitude $g \approx 25.9$ at $10\sigma$ for point sources \citep{ferrarese2012}.

At the distance of the Virgo Cluster, GCs and LMXBs can be treated as point sources, and the optical counterparts of AGN are dominated by accretion light from the nucleus, and can also be treated as point sources. Therefore, for the purpose of our study, we used the completeness limit for point sources, rather than the shallower limit for large extended sources like galaxies in Virgo. The surface luminosity of NGC 4472 \citep{mihos2013} is fainter than the sky brightness outside a radius of a few arcminutes (and about $3$ magnitudes fainter in the outer parts), and is not a limiting factor. We searched the NGVS point source catalog out to a radial distance that matched the extent of the X-ray data.

For the inner region, we used the X-ray sources within the four \textit{Hubble Space Telescope} fields described in \citet{kundu2002}.

\subsubsection{Active galactic nuclei and globular clusters}
\label{sect:agngc}

The vast majority of unobscured AGN and unidentified X-ray sources in \citet{lamassa2016} brighter than our X-ray completeness cutoff have $r \lesssim 25$. Almost all AGN have $g - r < 1$ \citep[e.g.,][]{richards2001}, and the few above this typically have high redshifts. AGN with redshifts larger than $3.5$ have X-ray fluxes fainter than our cutoff limit in $97\%$ of the \textit{Chandra Deep Field}-South sample \citep{vito2013}, which also used ACIS-I. This means that almost all AGN have $g < 26$, brighter than the NGVS detection limit for point sources. Systems dominated by galaxy starlight will have much higher ratios of optical to X-ray emission than unobscured AGN. Based on the GC luminosity function of NGC 4472 and other elliptical galaxies, almost all GCs at the distance of NGC 4472 have $g < 26$ \citep[e.g.,][]{rejkuba2001,kundu2002,kundu2007} and are expected to be listed in the NGVS catalog, and those that are fainter have low masses and are unlikely to host LMXBs. Therefore, both unobscured AGN and LMXBs located in GCs are optically bright enough for the NGVS to see.

\subsubsection{Low-mass X-ray binaries}

Generally, LMXBs in GCs and AGN have optical counterparts, whereas LMXBs in the field do not, but there are exceptions both ways. Here we considered optically bright field LMXBs, and we took a conservative approach when rejecting contaminating sources (sources that could be confused with field LMXBs).

The four X-ray sources with full-band ($0.5-7$ keV) X-ray flux $f_\mathrm{X} > 1 \times 10^{-13}$ erg s$^{-1}$ cm$^{-2}$ ($L_\mathrm{X} > 3 \times 10^{39}$ erg s$^{-1}$ if associated with NGC 4472) have clearly identified optical counterparts with $g < 23$, and the optical to X-ray luminosity ratios are between $0.3-1.2$. These four sources are most likely AGN (one is confirmed as such by SIMBAD Astronomical Database) or possibly LMXBs in GCs.

All other X-ray sources have $f_\mathrm{X} = 6.8 \times 10^{-15} - 7.4 \times 10^{-14}$ erg s$^{-1}$ cm$^{-2}$.
In this flux range, LMXBs with orbital periods below $10$ d have $g > 28$ according to the relation between optical and X-ray luminosity for a sample of NS and BH LMXBs by \citet{vanparadijs1994}. This is $1.5$ magnitude fainter than the $50\%$ completeness limit of the NGVS survey around NGC 4472 ($g_\mathrm{lim}$ = $26.5$, Stephen Gwyn, private communication). Therefore, virtually no LMXBs with orbital periods below $10$ d should have been detected in the NGVS catalog.
In the sample of LMXBs with known orbital periods in the catalog by \citet{liu2007}, $7$ out of $73$, about $10\%$, have orbital periods longer than $10$ d.
\footnote{In \citet{liu2007}, $60\%$ of LMXBs do not have a listed orbital period, but we found that the median HI extinction at their sky locations is $2.9$ times higher than for LMXBs with known orbital periods. We think that extinction and crowding near the Galactic Plane are the primary reasons why these systems do not have their orbital periods measured, rather than potential selection effects such as the difficulty in measuring very long and short orbital periods.}
In addition, LMXBs with $f_\mathrm{X} < 2 \times 10^{-14}$ erg s$^{-1}$ cm$^{-2}$ and orbital periods $< 30$ d also have $g > 28$. Even at brighter magnitudes, $g = 26-28$, many sources will be missing from the NGVS catalog given its $50\%$ completeness limit of $g_\mathrm{lim} = 26.5$.

Since about $10\%$ of LMXBs in our sample have orbital periods longer than $10$ d, enabling it to have $g < 28$, and furthermore at most a few tens of percent of these is actually detected by NGVS given the incompleteness of the catalog at such faint luminosities, we expect only a few percent of field LMXBs in our sample to have a detected optical counterpart. This may result in some field LMXBs to be removed from the sample, but this error is smaller than several other errors in our analysis. Also, if the actual detection limit for the faintest sources is shallower than the conservative estimate of $g = 28$, then the minimum orbital period LMXBs need to be optically bright enough to be detected becomes longer than $10$ d, and even fewer field LMXBs will be removed.

The vast majority of LMXBs are expected to be too optically faint for the NGVS to see. They have no optical counterpart in the NGVS data, except for spurious matches.
For the reasons described here and in Sect.~\ref{sect:agngc}, all X-ray sources with physically associated optical counterparts are assumed to be LMXBs in GCs or AGN, and are removed from the sample. X-ray sources without optical counterparts are assumed to be field LMXBs, our sources of interest. This way, the statistical number of field LMXBs in each annulus can be calculated, but because of spurious optical matches, we do not attempt to identify individual X-ray sources.

\subsubsection{Matching X-ray and optical sources}
\label{sect:halo:matching}

We used the analysis tool TOPCAT \citep{taylor2017} to cross-correlate the X-ray and optical source catalogs using a $1\arcsec$ matching radius. A test showed this matching radius to be optimal. With a lower matching radius, too many real counterparts are being missed, whereas using a larger radius gives almost no additional real counterparts, while errors become larger as a result of the strongly increasing number of chance matches. For our flux cutoff and corresponding photon count (about $16$ net counts), even at large off-axis angles a $1\arcsec$ matching radius is sufficient to find almost all real counterparts. We also verified this by observing that the density of optical sources between $1-4\arcsec$ away from X-ray sources at large off-axis angles is not significantly different from the density between $4 - 8\arcsec$. In Sect.~\ref{sect:halo:results} we included the effect of using different matching radii in the error estimates.

To estimate the rate of spurious associations, we also matched the X-ray and optical datasets after shifting by $5\arcsec$ in eight different directions (separated by $45\degr$ angles). This displacement distance is large enough to assure that X-ray sources do not match with their real optical counterparts, and at the same time small enough that the variation in the space density of GCs changes negligibly over this angular displacement. Consequently, any remaining matches must be chance coincidences. The average number of spurious matches over these displacements is subtracted from the number of matches in the real data, and the resulting figure is taken to be the number of X-ray sources with physically associated optical counterparts.

\subsubsection{Optically obscured active galactic nuclei}
\label{sect:halo:obscured}

AGN can be too optically faint to be detected in optical data if they are obscured \citep[e.g.,][]{risaliti1999,fabian1999,fiore2000,moretti2003}. Therefore, the remaining sample after optical elimination may still contain obscured AGN.

We estimated the fraction of AGN that are optically fainter than the NGVS limit using the Stripe~82 X-ray survey observations \citep[][VizieR Online Data Catalog]{lamassa2016}.
This survey targets blank fields, as it specifically avoids known targets such as nearby galaxies. It uses new and archival \textit{XMM-Newton} and archival \textit{Chandra} observations to observe $31.3$ deg$^{2}$ within the Sloan Digital Sky Survey (SDSS) Stripe~82 Legacy field \citep{frieman2008}, located within $1\fdg5$ of the celestial equator.
The flux limit in the full band ($0.5 - 7$ keV for \textit{Chandra}, $0.5 - 10$ keV for \textit{XMM-Newton}) is ${\sim}2.1 \times 10^{-15}$ erg s$^{-1}$ cm$^{-2}$. There are $5972$ X-ray sources brighter than our flux completeness cutoff. Of these, $25.8\%$ do not have an optical counterpart in SDSS data, down to about 25--26th magnitude. Since we only did optical matching, we did not consider the other multiwavelength data that are used by \citet{lamassa2016}.
As the SDSS photometry is about $2$ magnitudes shallower than the NGVS photometry, this percentage of obscured AGN is an upper limit for our purposes as it would probably be lower with deeper photometry.

We also looked at the fraction of AGN that are optically faint using two older and smaller studies that targeted other fields.

The Chandra Deep Survey \citep{hornschemeier2001} used ACIS-I to observe a field centered around the Hubble Deep Field--North field, and identified $82$ X-ray sources with a full-band ($0.5 - 8$ keV) X-ray sensitivity of ${\sim}3 \times 10^{-16}$ erg s$^{-1}$ cm$^{-2}$, and found counterparts to X-ray sources in \textit{I}-band photometry down to 26th magnitude, and \textit{R}-band to 25th magnitude. Nine X-ray sources have no optical or near-infrared counterpart, and the brightest of these is $8.2 \times 10^{-16}$ erg s$^{-1}$ cm$^{-2}$, almost an order of magnitude fainter than our completeness cutoff. The ten sources brighter than our cutoff are all detected in both the \textit{R}- and \textit{I}-bands, which are shallower than the NGVS photometry we used. X-ray sources in the blank field are expected to be AGN, and this result suggests that AGN in our data at X-ray fluxes brighter than our completeness cutoff are successfully removed from the sample using NGVS optical counterparts, although the sample of ten objects is small.

The Chandra Cal\'an-Yale Deep Extragalactic Research (CYDER) survey \citep{treister2005} used deep ACIS-I and ACIS-S observations of five fields at different parts of the sky, and identified $267$ X-ray sources, $119$ of which have full-band ($0.5 - 8$ keV) X-ray sensitivity brighter than our completeness cutoff. Two ACIS-S fields were directed at galaxy groups, and we excluded the four sources brighter than our cutoff that were identified as galaxies in the study. \citet{treister2005} looked for counterparts to X-ray sources in \textit{V}-band photometry down to $25.5 - 26.7$ magnitude ($5 \sigma$), and \textit{I}-band to $24.6 - 25.1$ ($5 \sigma$). These values may be slightly deeper than NGVS for some fields, depending on color and signal-to-noise conversions, but the CYDER X-ray exposure times are also longer. Of the remaining $115$ sources brighter than our cutoff, $81$ are detected in both bands, $15$ in a single band ($12$ in \textit{V}, three in \textit{I}), and $19$ are undetected. $19/115 = 16.5\%$ of X-ray sources have no optical/near-infrared counterpart, and again this fraction is an upper limit to the fraction using deeper optical data. Because some of the fields used are not blank but targeting compact groups of galaxies, it is also possible that some sources without optical counterpart are LMXBs, or perhaps lensed AGN in the background, implying a lower true fraction of obscured AGN.

Based on these studies we find that a fraction ${\sim}15\%$ of AGN in our data are expected to be obscured, and not be removed by optical matching.

\subsubsection{Sky angular density of active galactic nuclei}
\label{sect:halo:density}

We compared the number of AGN behind NGC $4472$ in our sample with AGN sky densities in other areas of the sky found by other studies. Due to large scale cosmological structures, however, AGN density varies significantly across the sky \citep{vikhlinin1995,cappi2001,cappelluti2009}. Therefore these values are not a good way to determine the AGN density behind NGC 4472, which is more reliably done by the optical-counterpart method described in Sect.~\ref{sect:agngc}.

To estimate the AGN sky density we used observations from the two surveys described in Sect.~\ref{sect:halo:obscured} that observed only blank fields.
The Stripe~82X survey found $5972$ X-ray sources brighter than our completeness cutoff, yielding a sky density of ${\sim}300$ deg$^{-2}$ based on \citet[][Fig.~3]{lamassa2016}. (Following \citet{lamassa2016}, we multiplied our cutoff flux by $1.21$ to convert the energy range from $0.5 - 7$ to $0.5 - 10$ keV).
The Chandra Deep Survey has ten sources (not including sources identified as stars) brighter than our cutoff in a sky area of approximately $17\arcmin \times 19\arcmin$, yielding a density of ${\sim}111$ deg$^{-2}$.
The CYDER \citep{treister2005} and ChaMP \citep{kim2007} surveys have higher sky densities, of $300 - 400$ deg$^{-2}$, than we found using the same completeness cutoff. In addition to cosmic variation, this can be due to these surveys making use of fields likely to contain foreground sources associated with the actual targets of the observations. Different models in converting count rate to X-ray flux can also play a role in variation in sky density.

\begin{figure}
\plotone{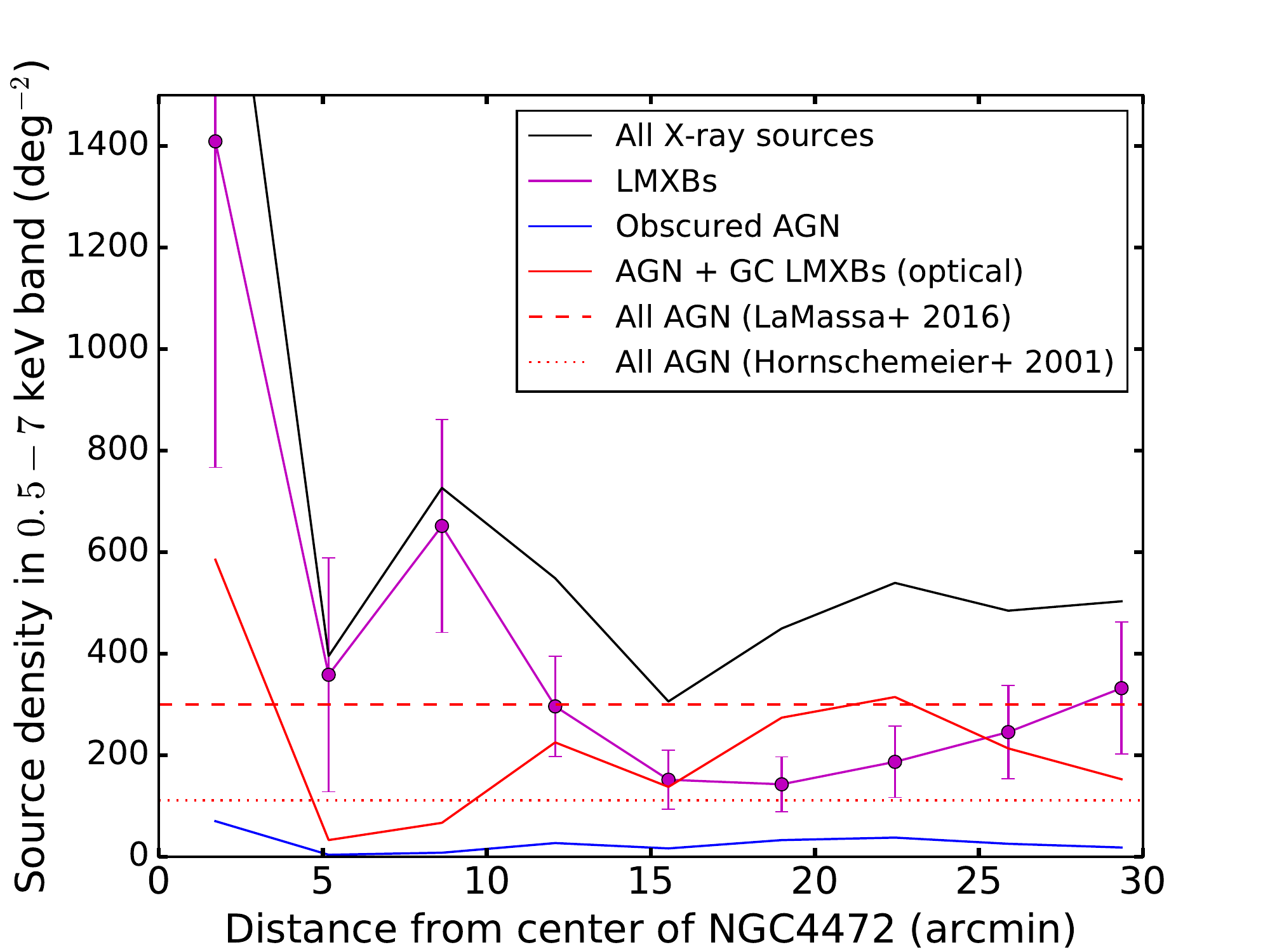} % Chandra_data/NGC4472/halo
\caption{Sky angular density of X-ray sources as a function of galactocentric radius. The projected distance is the semi-major axis of the elliptical annuli. The total density of X-ray sources (black curve) is the sum of field LMXBs (magenta curve), obscured AGN (blue curve), and optically-detected AGN plus LMXBs in GCs (i.e., the X-ray sources with optical counterpart -- solid red curve). The background AGN sky densities brighter than our completeness cutoff flux of $6.8 \times 10^{-15}$ erg s$^{-1}$ cm$^{-2}$, based on two studies of blank fields in different parts of the sky (not the Virgo Cluster), are shown as horizontal dashed and dotted red lines. The \citet{lamassa2016} figure is an upper limit, as it includes all X-ray sources in the sample. Error bars on the data points indicate $1\sigma$ confidence intervals. For clarity these are only shown on the field LMXB distribution.}
\label{fig:halo:density}
\end{figure}

Sky densities of important classes of sources are plotted against galactocentric radius in Fig.~\ref{fig:halo:density}. The semi-major axis distance of our \textit{Chandra} observations extends to almost $30\arcmin$, but the galactocentric radii of sources do not extend beyond $25\arcmin$. This is in part due to the distribution of LMXBs with azimuthal angle peaking near the east-west axis, about $55\degr$ away from the major axis of the light distribution. Depending on location, the density of AGN plus LMXBs in GCs is somewhat lower or similar to the density of AGN from \citet{lamassa2016}, which is by far the largest study of the two indicated by horizontal red lines. Overall the AGN densities are consistent with cosmic variation, especially since our field of view is only $0.4$ deg$^{2}$.
The density of AGN plus LMXBs in GCs in our sample at galactocentric radii of $4\arcmin - 10\arcmin$ is lower than the \citet{lamassa2016} value. This is partly due to small number statistics, because this part of the galaxy is covered only by the outer parts of the halo CCDs, which are aimed at ${\sim}15\arcmin$ from the Galactic center. Other complicating factors in the inner part of the galaxy are X-ray emission from diffuse gas, obscuration by dust, and crowding of X-ray and optical sources. Within $3\arcmin$ from the center of the galaxy, the high sky density of GCs will contribute significantly to the number of X-ray sources with optical match.

Because X-ray sources associated with GCs also have optical counterparts, it is not known precisely how many X-ray sources with optical counterparts are GCs, and how many are unobscured AGN. The number of obscured AGN follows from the (unknown) number of unobscured AGN combined with the fraction of AGN that are obscured (Sect.~\ref{sect:halo:obscured}), and cannot be calculated exactly. Since at most galactocentric radii the density of AGN by \citet{lamassa2016} is comparable to the unobscured AGN plus GC LMXB density in our result, probably most of the sources with optical counterparts are AGN rather than GCs. We conservatively assumed this to be $80\%$ in our calculations (see Sect.~\ref{sect:halo:discussion} for a justification).

The number density of obscured AGN that we estimated this way, assuming that $15\%$ of AGN are obscured even at the depth of the NGVS optical data (based on the surveys in Sect.~\ref{sect:halo:obscured}), is shown as the blue curve in Fig.~\ref{fig:halo:density}. There is a considerable uncertainty in the fraction of AGN that are obscured as a result of differences in the depths of optical and X-ray observations between studies. This uncertainty is taken into account in the error estimate of the results.
The obscured AGN density (blue curve) lies significantly below the LMXB density (magenta curve). This implies that obscured AGN are not likely to cause a significant contamination of the field LMXB density estimate.

\subsection{Surface brightness}
\label{sect:halo:surface}

The surface brightness of NGC 4472 decreases significantly from the center to the outer halo \citep[][Fig.~2, left panel]{mihos2013}. We considered the number of field LMXBs per unit stellar light, which approximates stellar mass. We calculated this ratio by dividing the total number of field LMXBs in the annulus by the luminosity of diffuse starlight within the annulus, derived from the surface brightness profile given in \citet{mihos2013}.

\section{Results}
\label{sect:halo:results}

We identified $344$ pointlike X-ray sources in NGC 4472, including $302$ in the halo in the five new \textit{Chandra} observations, as well as an additional $42$ sources in archival data from the inner part of the galaxy.
Our flux-limited complete sample consists of $174$ X-ray sources across all six observations, $86$ with an optical counterpart within $1\arcsec$ (either physically associated or chance superpositions), and $88$ without optical counterpart. Among the $174$ sources we estimated $91 \pm 15$ LMXBs, $74 \pm 14$ X-ray sources with a physically associated optical counterpart (LMXBs in GCs and AGN), and $9 \pm 5$ obscured AGN. $12 \pm 5$ X-ray sources with optical counterparts are false superpositions. All errors indicate $1\sigma$ confidence intervals. The errors in Fig.~\ref{fig:halo:density} are larger due to larger Poisson errors for the individual annuli compared to the entire sample \citep{kraft1991}.
The X-ray sources in our flux-limited sample have a lower limit of $6.8 \times 10^{-15}$ erg s$^{-1}$ cm$^{-2}$ for the $0.5-7$ keV band, or $2 \times 10^{38}$ erg s$^{-1}$ for sources located in NGC 4472, and therefore are more luminous than the Eddington limit for a $1.4\ M_{\sun}$ NS accreting hydrogen-rich matter.

\begin{figure}
\plotone{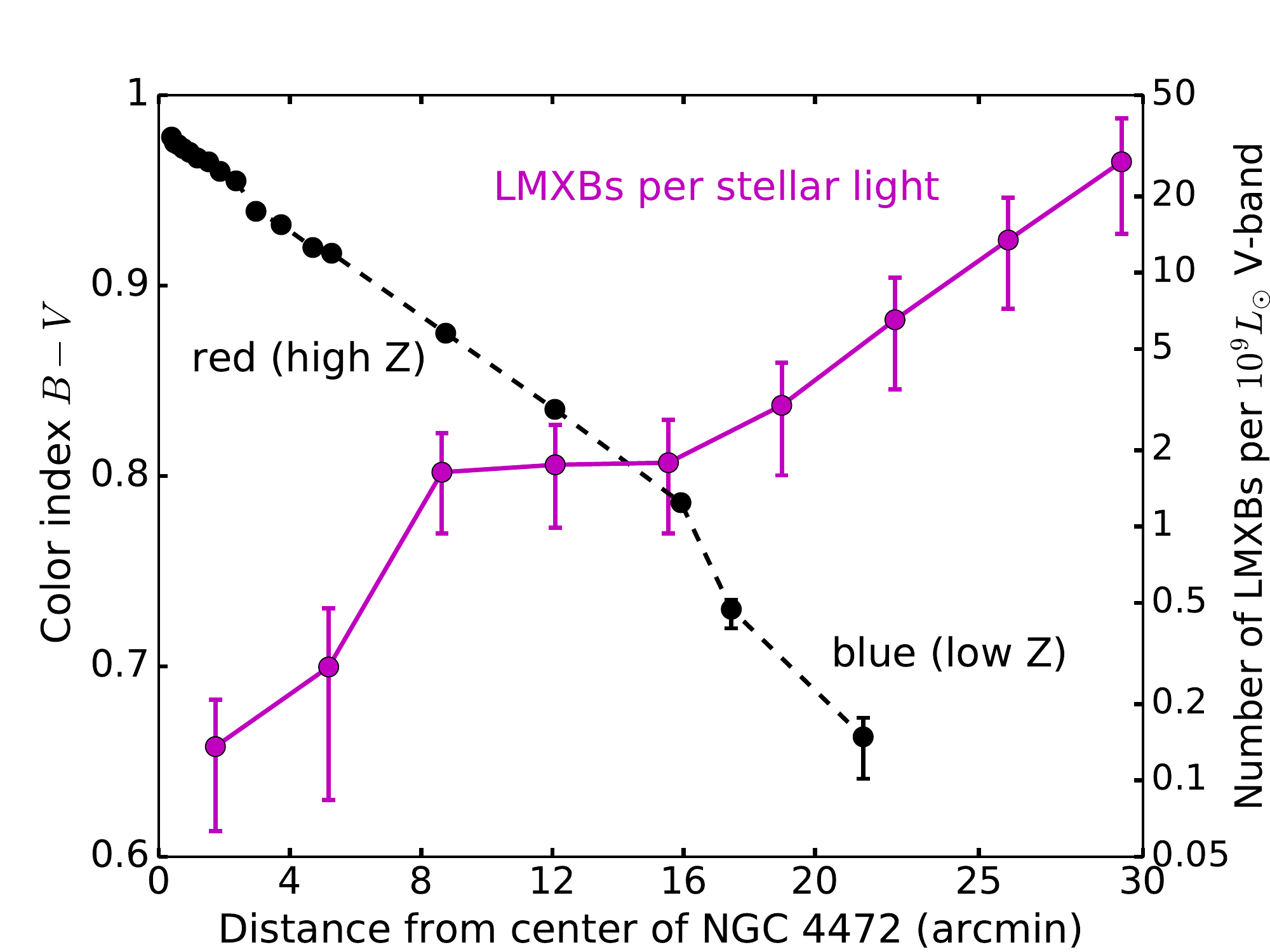} % Chandra_data/NGC4472/plot_halo
\caption{Radial distribution of field LMXBs with $L_\mathrm{X} > 2 \times 10^{38}$ erg s$^{-1}$ per unit stellar light (magenta circles connected by solid curve, axis on the right), after eliminating LMXBs in GCs and AGN. The projected distance is the semi-major axis of the elliptical annuli. This corresponds to the solid magenta curve in Fig.~\ref{fig:halo:density}. The color index (black circles connected by dashed curve, axis on the left) and surface brightness data are taken from \citet{mihos2013}. Error bars on the data points indicate $1\sigma$ confidence intervals.}
\label{fig:halo:main}
\end{figure}

Figure~\ref{fig:halo:main} shows the number of field LMXBs per unit \textit{V}-band luminosity (Sect.~\ref{sect:halo:surface}) for each annulus. The uncertainties in these field LMXB densities are a combination of Poisson errors regarding the number of sources per annulus, of the dependency of the result on the optical matching radius (Sect.~\ref{sect:halo:matching}), and of the fraction of AGN that are obscured (Sect.~\ref{sect:halo:obscured}).

Figure~\ref{fig:halo:main} also shows the $B - V$ color as a function of galactocentric radius \citep{mihos2013}. These colors indicate that metallicity decreases with galactocentric radius, and that the outer halo is very metal-poor. A~color of $B - V = 0.7$ corresponds to [Fe/H] $< -1$ \citep{mihos2013}.

There is an anticorrelation between color and the number of field LMXBs per unit stellar light. This implies an anticorrelation between number of field LMXBs and metallicity, in the sense that there are more field LMXBs per unit stellar light at bluer colors. This is the opposite of the metallicity effect seen in GCs.

A~$\chi^{2}$ test comparing the number of field LMXBs per unit stellar light to a constant distribution equal to its weighted average (with weight factors of $\mathrm{error}^{-2}$) yields $\chi^{2}/\mathrm{dof} = 29.9/8 = 3.7$ for $9$ bins, corresponding to a $p$ value of $2 \times 10^{-4}$.
The probability that the field LMXB density per unit stellar light is consistent with a constant value based on a Kolmogorov-Smirnov test is $p = 2.0 \times 10^{-5}$. This test yields a statistic $D = 0.626$ for the cumulative distribution of \textit{V}-band light in each annulus versus the distribution of field LMXBs in each annulus, when taking $50$ bins for the field LMXB distribution (instead of $9$) to obtain a more significant result, and $19$ bins for the flux distribution \citep[][Fig.~2, left panel]{mihos2013}.
Both tests indicate that the number of field LMXBs per unit stellar light is significantly increasing with galactocentric radius.

\begin{figure}
\plotone{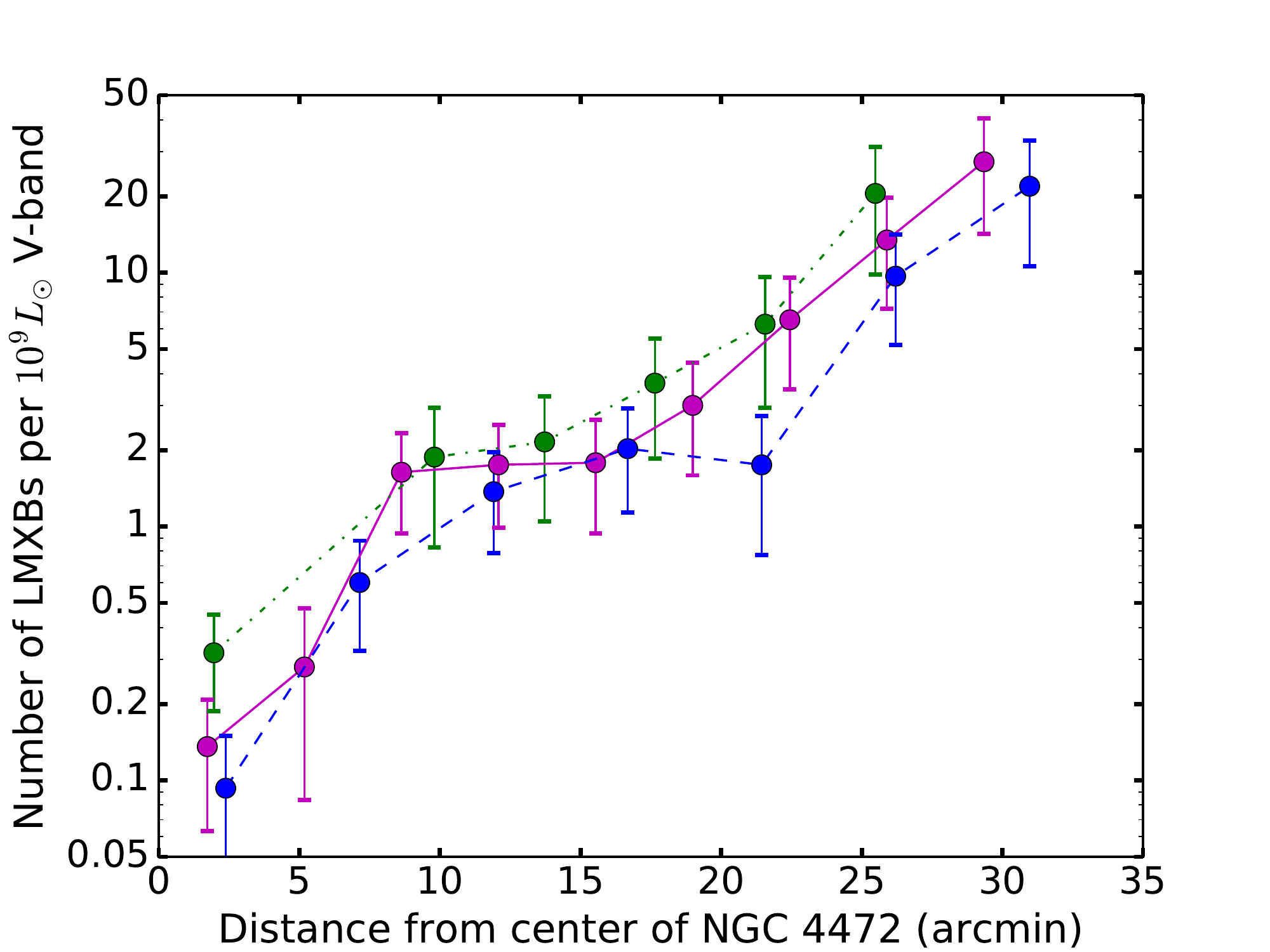} % Chandra_data/NGC4472/plot_halo
\caption{Radial distribution of field LMXBs per unit stellar light after eliminating LMXBs in GCs and AGN. The solid magenta curve is identical to the one in Fig.~\ref{fig:halo:main}. The dash-dotted green and dashed blue curves represent samples with lower ($5\farcm4$) and higher ($12\arcmin$) off-axis angle cutoffs, listed in Table~\ref{tab:stat}. The projected distance is the semi-major axis of the elliptical annuli. Error bars on the data points indicate $1\sigma$ confidence intervals.}
\label{fig:halo:samples}
\end{figure}

Other choices for the off-axis angle cutoff lead to different complete samples. In the case of a smaller cutoff angle, fainter sources are included, but sources that are farthest off-axis are excluded (Fig.~\ref{fig:halo:oaa}). This improves the quality of the sample at galactocentric radii that are relatively close to those of the aimpoints of the halo observations, but comes at the cost of losing information at large galactocentric radii. When the off-axis angle cutoff differs from $9\farcm6$, the number of sources retained in the sample is lower (per Fig.~\ref{fig:halo:options}).
Figure~\ref{fig:halo:samples} shows the same radial distribution of field LMXBs as in Fig.~\ref{fig:halo:main} (magenta curve), but now compares it with two more flux-limited samples; one with a lower off-axis angle cutoff and a correspondingly lower flux limit (dash-dotted green), and a sample with a higher off-axis angle cutoff and a higher flux limit (dashed blue). Table~\ref{tab:stat} lists the off-axis angles and flux limits for all three samples, as well as the statistics for the anticorrelation between color and field LMXBs per unit light.
The main results are similar; the number of field LMXBs per unit light still increases significantly with galactocentric radius. The statistical tests for a lower off-axis cutoff of $5\farcm4$ are less significant than those for our standard sample with an off-axis cutoff of $9\farcm6$. This is partly due to a lack of coverage of galactocentric radii in the range of $4-8\arcmin$, and partly due to a smaller overall sample size.

\begin{table}
\caption{Parameters for three samples of LMXBs, and statistical test results for the anticorrelation between color and number of field LMXBs per unit stellar light. OAA stands for off-axis angle, \textit{F} is the lower flux limit in units of $10^{-15}$ erg s$^{-1}$ cm$^{-2}$. See text for more details.}
\label{tab:stat}
\begin{tabular}{ccccccc}
\hline
OAA & \textit{F} & $\chi^{2}/\mathrm{dof}$ & \textit{p} ($\chi^{2}$) & KS & \textit{p} (KS) \\
\hline
$5\farcm4$  & $3.0$ & $16.2/5 = 3.2$ & $6 \times 10^{-3}$ & $0.525$ & $7 \times 10^{-4}$ \\
$9\farcm6$  & $6.8$ & $29.9/8 = 3.7$ & $2 \times 10^{-4}$ & $0.626$ & $2 \times 10^{-5}$ \\
$12\arcmin$ & $9.0$ & $23.3/6 = 3.4$ & $7 \times 10^{-4}$ & $0.633$ & $2 \times 10^{-5}$ \\
\hline
\end{tabular}
\end{table}

The near-infrared $K$ band is a better indicator for mass than optical light \citep[e.g.,][]{bell2001}. Using the relation between $B - V$ and $V - K$ in \citet[][Table 5]{pecaut2013}, it shows that $V - K$ decreases from the inner to the outer parts of the galaxy. Therefore, the surface brightness in the $K$ band falls off steeper with radius than the $V$-band surface brightness. This means that the anti-correlation result will become more significant when expressed in units of $K$-band light rather than $V$-band light, as we did in Figs.~\ref{fig:halo:main} and \ref{fig:halo:samples}.

The number of field LMXBs $N$ per unit stellar light for our standard sample with off-axis angle cutoff $9.6\arcmin$ can be fitted by $\log_{10} (N / L_{V \sun}) = {(0.076 \pm 0.007) r} - {(10.01 \pm 0.13)}$, with $L_{V\sun}$ the solar \textit{V}-band luminosity and $r$ the galactocentric radius in arcmin. The relation between $N$ and color is fitted by $\log_{10} (N / L_{V \sun}) = {(-4.7 \pm 0.5)} {(B-V)} - {(5.3 \pm 0.4)}$. The uncertainties in these fits indicate $1\sigma$ statistical errors.
The $1\sigma$ relative error on the slope in the $\log_{10} (N / L_{V \sun})$ versus $r$ relation of $9\%$ is another indication that this slope being positive is significant. The slope of the $\log_{10} (N / L_{V \sun})$ versus $B-V$ relation being negative is also significant.

\section{Discussion}
\label{sect:halo:discussion}

Our assumptions in Sect.~\ref{sect:halo:density} that $80\%$ of sources with an optical counterpart are AGN and $15\%$ of AGN are obscured are probably conservative. Lowering either number decreases the estimated number of obscured AGN (blue curve in Fig.~\ref{fig:halo:density}) and increases the estimated number of field LMXBs (magenta curve), and makes the result that the number of field LMXBs per unit stellar light increases with galactocentric radius statistically more significant. Assuming the most conservative percentage of $100\%$ instead of $80\%$ (i.e., assuming there are no GC X-ray sources) produces a result that is almost equally significant, because the corresponding increase in the absolute number of obscured AGN ($15\%$ of all AGN) is small. Even in the extreme case in which $100\%$ of optical sources are AGN, and $75\%$ of AGN are obscured instead of $15\%$ (which is certainly not realistic for our X-ray flux limit combined with the deep optical NGVS observations, according to the AGN studies cited in Sect.~\ref{sect:halo:obscured}) the KS test still gives $1.0 \times 10^{-3}$ instead of $2.0 \times 10^{-5}$.

If LMXBs are predominantly formed at the galactocentric radii where they are found today, their high number density per unit light in the metal-poor outer halo would require a reversed metallicity effect, opposite from that seen in GCs.
\citet{mihos2013} considered it very unlikely that the steep color gradient in NGC 4472 is a consequence of a decrease in stellar ages (rather than metallicity) with galactocentric radius because it would require a very large and relatively recent merger event, which they found to be inconsistent with the observed accretion features in the galaxy. Merger events of any magnitude can initiate local star formation and potentially lead to an excess of LMXBs at larger galactocentric radii, but we did not find clear evidence for an azimuthal pattern in the LMXB distribution.
Natal kicks may explain the observed field LMXB distribution, if most binaries formed in the inner galaxy where most of the stellar mass is, and many the surviving systems reached the outer halo after receiving kicks. However, the required kicks may be larger than those inferred from Galactic radio pulsars \citep{hobbs2005}.
Many LMXBs, even those from the higher (non-electron capture) peak in the kick velocity distribution \citep[e.g.,][]{arzoumanian2002}, could have been retained in the galaxy due to the deep potential of NGC 4472, making this massive galaxy a good test location for natal kicks on NSs and BHs. Since the lower limit on the flux of $6.8 \times 10^{-15}$ erg s$^{-1}$ cm$^{-2}$ in our standard sample corresponds to the Eddington limit for NSs, many of the LMXBs in the complete sample may host a BH accretor.

Another possibility is that some field LMXBs originate from GCs from which they were dynamically ejected, or which were evaporated \citep{kundu2002,kundu2007}. However, \citet{juett2005} and \citet{irwin2005} found that a significant number of LMXBs in elliptical galaxies must have formed in the field rather than in GCs, and \citet{kim2006} did not find evidence either for or against the idea that all LMXBs are formed in GCs. Also, kick velocities complicate comparison of radial profiles of GCs and field LMXBs, which in addition are usually not available for the outer halo, which we studied.

\subsection{Previous studies of extragalactic halos}

Several studies also found an increase in the number of LMXBs per unit stellar light towards larger galactocentric radii, although these did not extend as far out as our data.

In the halo of M104, \citet{li2010} found an excess of X-ray sources compared to starlight between galactocentric radii of $4\arcmin - 9\arcmin$ \citep[$10 - 24$ kpc, based on a distance of $9.0$ Mpc,][]{spitler2006}, double the number expected from background AGN. As possible explanations, the authors considered ejection from inner region of the galaxy as a result of supernova kicks, association with GCs, formation following a recent galaxy merger, or a strong overdensity in the AGN background. \citet{li2010} subsequently also found excesses in some other galaxies.

\citet{zhang2013} found an excess compared to stellar light in the outskirts of $20$ galaxies, at radii of ${\sim}2\arcmin - 9\arcmin$, and concluded that both galaxy mass and LMXBs in GCs contribute.
In NGC 4365 the authors were able to eliminate LMXBs in GCs using optical counterparts, and found an excess of field LMXBs between $2\farcm5 - 6\farcm4$ ($15 - 38$ kpc).

\citet{mineo2014} also found an excess of LMXBs (that could be either field or GC sources) compared to \textit{V}-band stellar light up to $7\farcm5$ (${\sim}37$ kpc) from the center in the outskirts of NGC 4649 (M60), which has a distance of $16.8$ Mpc \citep{tonry2001}. They suggested galaxy interaction causing a rejuvenated field LMXB population as explanation, but LMXBs in GCs may also play a role due to lack of deep GC data at large radii. \citet{mineo2014} considered natal kicks to be unlikely as main cause due to the shape of their source distribution, which spikes near $6\farcm5 - 7\arcmin$ from the center of NGC 4649.

\section{Summary}

We obtained \textit{Chandra} X-ray observations of the outer halo of NGC 4472, and combined with existing X-ray and ground-based optical data, we showed that the number of field LMXBs per unit stellar light increases significantly from the nucleus towards the outer halo. Migration of binaries caused by natal kicks onto NSs and BHs can possibly explain this in part, but the variation in metallicity across the halo may also play a role.

\acknowledgments

The authors thank Stephanie LaMassa for providing data from the Stripe~82X survey, Stephen Gwyn and the Next Generation Virgo Cluster Survey team for providing optical data, and Ezequiel Treister, Jay Strader, Gregory Sivakoff, Vlad Tudor, and the anonymous referees for useful discussion and comments.
Support for this work was provided by NASA through Chandra Award Numbers GO4-15034 and GO5-16084 issued by the Chandra X-ray Center (CXC), which is operated by the Smithsonian Astrophysical Observatory for and on behalf of NASA under contract NAS8-03060.
The scientific results reported in this article are based on observations made by the \textit{Chandra X-ray Observatory}, and on data obtained from the Chandra Data Archive.
This research has made use of software provided by the CXC in the application package CIAO, as well as NASA's Astrophysics Data System Bibliographic Services.

\vspace{5mm}
\facility{CXO}
\software{CIAO \citep{fruscione2006}, TOPCAT \citep{taylor2017}}

\bibliography{lennart_refs}

\end{document}